\documentclass{ws-procs9x6}
\newcommand{\zpr}{\mbox{$Z'$}}
\newcommand{\mzp}{\mbox{$M_{Z'}$}}
\newcommand{\upr}{\mbox{$U(1)'$}}

\begin{document}
\title{An NMSSM Without Domain Walls
\footnote{\uppercase{T}alk presented by \uppercase{Bob McElrath}
at {\it \uppercase{SUSY} 2003:
\uppercase{S}upersymmetry in the \uppercase{D}esert}\/, 
held at the \uppercase{U}niversity of \uppercase{A}rizona,
\uppercase{T}ucson, \uppercase{AZ}, \uppercase{J}une 5-10, 2003.
\uppercase{T}o appear in the \uppercase{P}roceedings.}
}


\author{Tao Han}

\address{Department of Physics \\
University of Wisconsin\\ 
Madison, WI 53706 USA\\ 
E-mail: than@pheno.physics.wisc.edu}

\author{Paul Langacker}

\address{Department of Physics and Astronomy\\ 
University of Pennsylvania\\
Philadelphia, PA 19104 USA\\ 
E-mail: pgl@electroweak.hep.upenn.edu}  

\author{Bob McElrath}

\address{Department of Physics \\
University of California\\ 
Davis, CA 95616 USA\\ 
E-mail: mcelrath@physics.ucdavis.edu}


\maketitle

\abstracts{
    We consider the Higgs sector in an extension of the MSSM involving an extra
    \upr \ gauge symmetry and SM singlet \upr charged scalars, in which the
    effective $\mu$ parameter is decoupled from the \zpr \ mass.  There are
    large mixings between Higgs doublets and singlets, significantly affecting
    the Higgs spectrum, production cross sections, decay modes, exclusion
    limits, and allowed parameters ranges. Scalars considerably lighter than 114
    GeV are allowed, and the range $\tan \beta \sim 1$ is both allowed and
    theoretically favored. We concentrate on the lighter (least model
    dependent) Higgs particles with significant $SU(2)$-doublet components to
    their wave functions, for the case of no explicit $CP$ violation in the
    Higgs sector.  We consider their spectra, including the dominant radiative
    corrections to their masses in the large $\tilde t$ mass limit; production
    cross sections and exclusion limits at LEP and a future linear collider;
    and decay properties.\cite{us}
}

\section{Introduction}

The superpotential for the MSSM contains the supersymmetric mass term $\mu H_2
H_1$.  The minimization condition for the MSSM scalar potential relates $\mu$ to
$M_Z$ and soft SUSY breaking parameters.  However, $\mu$ is an input-scale
parameter and therefore should have a value ${\mathcal O}(M_{Pl})$ or ${\mathcal
O}(M_{GUT})$.  This has led to a widespread belief that the MSSM must be
extended at very high energies to include a mechanism which relates $\mu$ to the
SUSY breaking mechanism.\cite{muprob}  The simplest extension\cite{nmssm} adds a
single scalar singlet with the superpotential $\lambda S H_u H_d + k S^3$, which
is able to solve the $\mu$ problem at the expense of requiring a $Z_3$ discrete
symmetry to forbid a $\mu_S S^2$ term which would re-introduce the $\mu$
problem.  It has been criticized because it would create domain walls in the
early universe.  Additionally, in many string constructions (e.g., heterotic and
intersecting brane) superpotential terms are typically off-diagonal in the
fields, which disallows the $k S^3$ term.  Thus we are led to consider larger
models that may be derived from string constructions involving a non-anomalous
\upr.

\upr\ models involve a standard model singlet field $S$ which yields an
effective $\mu$ parameter $h \langle S \rangle$, where the superpotential
includes the term $h SH_2H_1$, thus solving the $\mu$ problem\cite{mugauge}.
$S$ will be charged under the \upr, so that its expectation
value also gives mass to the new \zpr \ gauge boson. The extended gauge symmetry
forbids an elementary $\mu$ term as well as terms like $S^n$ in the
superpotential (the role of the $S^3$ term in generating quartic terms in the
potential is played by $D$ terms and possibly off-diagonal superpotential terms
involving additional standard model singlets). 

In this talk I explore the extended Higgs sector in a particular \upr \ model
using a bottom-up approach involving several standard model singlet
fields\cite{ell}.  This model has the advantage that it somewhat decouples the
effective $\mu$ parameter from the \zpr \ mass, and leads naturally to a
sufficiently heavy \zpr.  We make the assumption that any fermionic matter
necessary to cancel anomalies also receives a mass at the \upr\ breaking scale
and is decoupled from physics that will be visible at the LHC.  We expect that
the generic features will be representative of a wider class of constructions.  

\section{The Model}

The model we consider, first introduced in~\cite{ell}, has the superpotential:
\begin{equation}
  W = h S H_2 H_1 + \lambda S_1 S_2 S_3 + W_{\rm MSSM} \label{wmodel}
\end{equation}
$S$, $S_1$, $S_2$, and $S_3$ are standard model singlets, but are charged under
an extra $U(1)^\prime$ gauge symmetry.  The model is such that the potential has
an $F$ and $D$-flat direction in the limit $\lambda \rightarrow 0$, allowing a
large (TeV scale ) \zpr \ mass for small $\lambda$.  The use of an $S$ field
different from the $S_i$ in the first term allows a decoupling of \mzp \ from
the effective $\mu$.

The superpotential $W$ leads to the $F$-term scalar potential:
\begin{eqnarray}
  \label{VF}
  \nonumber    
    V_F = & h^2
      \left(|H_2|^2 |H_1|^2 + |S|^2 |H_2|^2 + |S|^2 |H_1|^2\right) \\
    + & \lambda^2 
      \left( |S_1|^2 |S_2|^2 + |S_2|^2 |S_3|^2 + |S_3|^2 |S_1|^2 \right).
\end{eqnarray}
The $D$-term potential is:
\begin{eqnarray}
  \label{VD}
  V_D &=& \frac{{G^2}}{8} \left(|H_2|^2 - |H_1|^2\right)^2 \nonumber\\&
      + & \frac{1}{2} g_{Z'}^2\left(Q_S |S|^2 + Q_{H_1} 
      |H_1|^2 + Q_{H_2} |H_2|^2 + \sum_{i=1}^3 Q_{S_i}
      |S_i|^2\right)^2 ~,~\,
\end{eqnarray}  
where $G^2=g_1^{2} +g_2^2=g_2^2/\cos^2 \theta_W$. $g_1, g_2$,  and $g_{Z'}$ are
the coupling constants for $U(1), SU(2)$ and $U(1)^{\prime}$, respectively, and
$\theta_W$ is the weak angle.  $Q_{\phi}$ is the $U(1)^{\prime}$ charge of the
field $\phi$. We will take $g_{Z'} \sim \sqrt{5/3} g_1$ (motivated by gauge
unification) for definiteness.

We do not specify a SUSY breaking mechanism but rather parameterize the breaking
with the soft terms
\begin{eqnarray}
\label{vsoft}
    V_{\rm soft} &=& m_{H_1}^2 |H_1|^2 + m_{H_2}^2 |H_2|^2 + m_S^2 |S|^2 +
       \sum_{i=1}^3 m_{S_i}^2 |S_i|^2
       \nonumber\\&
        - & (A_h h S H_1 H_2 + A_{\lambda} \lambda S_1 S_2 S_3 + {\rm H. C.})
    \nonumber \\
&+& (m_{SS_1}^2 S S_1 + m_{SS_2}^2 S S_2 + {\rm H. C.})
\end{eqnarray} 
The last two terms are necessary to break two unwanted global $U(1)$ symmetries,
and require $Q _{S_1}=Q _{S_2}=-Q_S$.  The potential $V=V_F+V_D +V_{soft}$ was
studied in~\cite{ell}, where it was shown that for appropriate parameter ranges
it is free of unwanted runaway directions and has an appropriate minimum.

\section{Higgs sector and electroweak symmetry breaking}
\label{EWSB}

The Higgs sector for this model contains 6 CP-even scalars and 4 CP-odd
scalars, which we label $H_1 ... H_6$ and $A_1 ... A_4$, respectively, in order
of increasing mass.  We compute the six CP-even scalar masses including the
dominant 1-loop contribution coming from the top/stop loop.  We scan over
vacuum expectation values such that the three singlets $S_1$, $S_2$, and $S_3$
typically have larger vevs than the other three fields.  

\begin{figure}[tb]
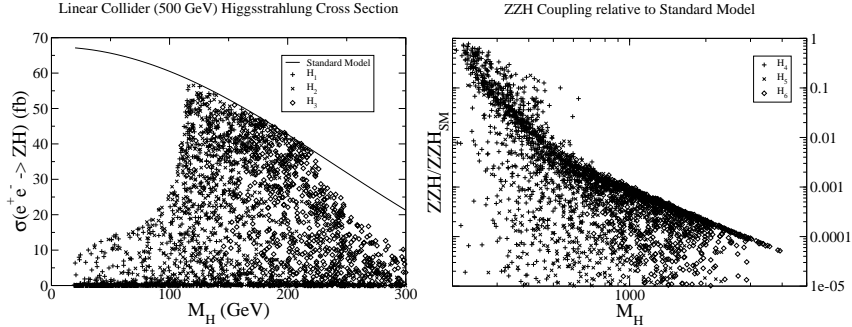

    \includegraphics[scale=0.22]{xsect_ZH_123_bw.eps}
    \includegraphics[scale=0.22]{ZZHcoupling_mh_456_bw.eps}
    \caption {Higgsstrahlung cross section for a 500 GeV linear
    collider (left) and the $ZZH$ coupling relative to the Standard Model for the
    3 heaviest CP-even states (right).}

    \label{Higgsstrahlung}
\end{figure}

A scan of parameter space indicates that there is a significant region with
light $H$ and/or light $A$, able to escape experimental detection at LEP2
because of its small mixing with the MSSM Higgs doublets.  In
Fig.\ref{Higgsstrahlung} we present the cross section for the Higgsstrahlung
process $e^+e^- \rightarrow ZH$ for the lightest 3 CP-even states, and the $ZZH$
coupling of the heaviest states relative to the Standard Model.  One can see
that $20 {\rm GeV}<m_{H_1} < 200 {\rm GeV}$ is allowed, and the heaviest states
are decoupled from the Z.

Decays of the Higgses in this model can be very unlike the MSSM, with $H
\rightarrow A_1 A_1$ and $H \rightarrow \chi^0 \chi^0$ dominating if they are
kinematically allowed due to new \upr\ D-term and singlet F-term contributions.
In addition $H \rightarrow W^+ W^-$ can be kinematically allowed.
Further phenomenological analysis including $HA$ production, $H$ and $A$ decay
modes, and gauginos will be presented in Ref.\cite{us}.

\section*{Acknowledgments}
This work was supported in part by DOE grant
DE--FG02--95ER--40896, in part by the University of Wisconsin
Research Committee with funds granted by the Wisconsin Alumni Research
Foundation, and in part by DOE grant DOE-EY-76-02-3071.


\end{document}